\documentclass[reprint,canadian,english,jcp,notitlepage,citeautoscript,longbibliography]{revtex4-1}
\usepackage[T1]{fontenc}
\usepackage[latin9]{inputenc}
\usepackage{babel}
\usepackage{mathrsfs}
\usepackage{amsmath}
\usepackage{amssymb}
\usepackage{graphicx}
\usepackage[unicode=true,pdfusetitle,
 bookmarks=true,bookmarksnumbered=false,bookmarksopen=false,
 breaklinks=false,pdfborder={0 0 1},backref=false,colorlinks=false]
 {hyperref}
\hypersetup{
 pdfborderstyle=}

\makeatletter

\providecommand{\tabularnewline}{\\}
\newcommand{\lyxdot}{.}

\usepackage{babel}
\usepackage{braket}

\makeatother

\begin{document}
\selectlanguage{canadian}%
\global\long\def\d{\mathrm{d}}%

\global\long\def\abs#1{|#1|}%

\global\long\def\braket#1#2{\Braket{#1|#2}}%

\global\long\def\bra#1{\Bra{#1}}%

\global\long\def\ket#1{\Ket{#1}}%

\global\long\def\kket#1{|#1\rangle\kern-0.25em  \rangle}%

\global\long\def\bbra#1{\langle\kern-0.25em  \langle#1|}%

\global\long\def\bbrakket#1#2{\langle\kern-0.25em  \langle#1|#2\rangle\kern-0.25em  \rangle}%

\global\long\def\alg{\text{UF}^{2}}%

\global\long\def\cvec#1{\mathbf{#1}}%

\renewcommand{\Im}{\text{Im}}

\providecommand{\Tr}{\mathrm{Tr}}
\title{Automatic Feynman diagram generation for nonlinear optical spectroscopies}
\author{Peter A.~Rose }
\affiliation{Department of Physics, University of Ottawa, Ottawa, ON, K1N 6N5,
Canada}
\author{Jacob J.~Krich}
\affiliation{Department of Physics, University of Ottawa, Ottawa, ON, K1N 6N5,
Canada}
\affiliation{School of Electrical Engineering and Computer Science, University
of Ottawa, Ottawa, ON, K1N 6N5, Canada}
\begin{abstract}
Perturbative nonlinear optical spectroscopies are powerful methods
to understand the dynamics of excitonic and other condensed phase
systems. Feynman diagrams have long provided the essential tool to
understand and interpret experimental spectra and to organize the
calculation of spectra for model systems. When optical pulses are
strictly time ordered, only a small number of diagrams contribute,
but in many experiments pulse-overlap effects are important for interpreting
results. When pulses overlap, the number of contributing diagrams
can increase rapidly, especially with higher order spectroscopies,
and human error is especially likely when attempting to write down
all of the diagrams. We present an automated Diagram Generator (DG)
that generates all of the Feynman diagrams needed to calculate any
$n^{\text{th}}$-order spectroscopic signal. We characterize all perturbative
nonlinear spectroscopies by their associated phase-discrimination
condition as well as the time intervals where pulse amplitudes are
nonzero. Although the DG can be used to automate impulsive calculations,
its greatest strength lies in automating finite pulse calculations
where pulse overlaps are important. We consider third-order transient
absorption spectroscopy and fifth-order exciton-exciton interaction
2D (EEI2D) spectroscopy, which are respectively described by 6 or
7 diagrams in the impulsive limit but 16 or 240 diagrams, respectively,
when pulses overlap. The DG allows users to automatically include
all relevant diagrams at relatively low computational cost, since
the extra diagrams are only generated for the inter-pulse delays where
they are relevant. For EEI2D spectroscopy, we show the important effects
of including the overlap diagrams.
\end{abstract}
\maketitle

\section{Introduction}

Nonlinear optical spectroscopies (NLOS) are powerful tools for elucidating
excited state dynamics of a wide variety of condensed phase systems
and have been particularly important for determining the evolution
of excitonic systems \citep{Zhang1998,Brixner2005,Cho2005,Abramavicius2009,Panitchayangkoon2011,Christensson2012,Fetherolf2017,Jang2018,Cruz2019,Kiessling2020,Maly2020,Suess2020}.
Interpretation of data-rich NLOS signals is centered on Feynman diagrams,
which conveniently graphically summarize  time-dependent perturbation
theory contributions to the signals \citep{Mukamel1999}. These diagrams
give a visual understanding of what types of excited- and ground-state
dynamics and/or coherences are probed and can be straightforwardly
turned into calculations of contributions to signals. In many cases,
diagrams can be determined to give zero contribution without a complicated
calculation, making theoretical interpretation easier and calculation
less expensive.

When pulse durations are shorter than system dynamics -- the impulsive
limit -- the number of contributing diagrams is often small, especially
for third-order spectroscopies. It has been easiest to build intuition
about the experimental signatures of particular excitonic processes
by considering the impulsive limit \citep{Brixner2005,Cho2005,Yang2007,Huo2011,Panitchayangkoon2011,Perdomo-Ortiz2012,Christensson2012,Dostal2012,Karki2014,Fetherolf2017,Provazza2018,Dostal2018,Maly2018}.
When pulse durations are similar to the timescale of system evolution,
however, the effects of pulse overlaps -- where the tail of a nominally
earlier pulse arrives after the beginning of a nominally later pulse
-- become important to accurately model and understand experimental
results \citep{Faeder1999,Belabas2004,Yetzbacher2007,Tekavec2010,Yuen-Zhou2012,Tiwari2013,Li2013,Leng2016,Perlik2017,Smallwood2017,Do2017,Palecek2019,Anda2020}.
Considering such processes requires many more contributing diagrams.
Higher-order spectroscopies, both in the impulsive and pulse-overlap
limits, also involve rapidly increasing numbers of contributing diagrams.
Human error and fatigue in determining these diagrams accurately become
increasingly likely as their numbers proliferate.

We present here an automated Feynman Diagram Generator (DG), which
allows convenient, fast, and accurate determination of the diagrams
contributing to a particular spectroscopy. We describe all perturbative
nonlinear spectroscopies by their associated phase-discrimination
conditions. These conditions allow description of both non-colinear
phase-matching and colinear phase-cycling experiments \citep{Mukamel1999,Tekavec2007,Cho2009,PeterHamm2011,Perdomo-Ortiz2012,Nardin2013,Bakulin2016,Tiwari2018,Maly2018}.
Users input those conditions and the time intervals in which their
pulses are nonzero, and the DG produces a list of contributing diagrams.
That list can be passed directly to calculation engines to predict
the associated spectroscopic signals or can be drawn in the standard
diagrammatic format for review. The DG is part of a set of NLOS tools
called the Ultrafast Spectroscopy Suite (UFSS), which also contains
methods to generate Hamiltonians or Liouvillians for vibronic systems
and to calculate the contributions from each diagram \citep{Rose2019,Rose2020b},
but its output diagrams can be used with other computational tools
or analytic methods to calculate spectroscopic signals \citep{Engel1991,Beck2000,Belabas2004,Renziehausen2009,Tanimura2012,Johansson2012,Yuen-Zhou2014,Albert2015,Perlik2017,Yan2017,Smallwood2017,Fetherolf2017,Provazza2018,Ke2018a,quantarhei}.
The DG is free and open-source software written in Python, available
for download from \href{https://github.com/peterarose/ufss}{github}.

The DG (and the larger UFSS) is designed to make simple the inclusion
of the effects of finite-duration pulses in NLOS. While modeling often
considers the impulsive limit, pulse-overlap effects can dominate
third-order signals such as two-dimensional photon echo (2DPE) spectroscopy,
even outside of what is commonly thought of as the pulse-overlap window
\citep{Palecek2019}. 2DPE is usually calculated using 6 time-ordered
diagrams (3 for the rephasing and 3 for the non-rephasing pathway).
When pulses overlap, up to 16 diagrams (shown in Fig.~\ref{fig:All-rephasing})
contribute to the signal \citep{Jonas2003}. These 16 diagrams have
been written down by hand and used for calculations in the past. 

In higher-order methods, such as fifth-order exciton-exciton interaction
2D spectroscopy (EEI2D), there are 7 Feynman diagrams in the impulsive
limit \citep{Dostal2018}. We show that outside of the impulsive limit,
EEI2D requires up to 240 Feynman diagrams. This number of diagrams
is too large for generation by hand, and as a result we do not believe
a calculation with all of these diagrams has previously been attempted
in the perturbative limit. References ~\onlinecite{Do2018,Suess2020}
consider finite pulse effects from time-ordered diagrams for several
$5^{th}$-order spectroscopies. We use EEI2D as our key example of
the application of automated diagram generation, showing both 1) that
these extra overlap diagrams can be generated and their contributions
calculated with surprisingly small extra computational cost and 2)
that finite pulse effects arising from non-time-ordered diagrams can
provide significant modifications to a sample EEI2D spectra, which
can be important for interpretation of experimental results. 

The DG automatically creates all of the diagrams that satisfy a given
phase-discrimination condition. Since the DG is computationally inexpensive
compared to evaluating the contribution of each diagram, in UFSS the
DG is run for each set of desired pulse delays. It determines whether
pulses overlap, and thus whether overlap diagrams contribute, allowing
computation of only causal diagrams for each set of pulse delays.
This determination provides a significant computational time savings,
since the large number of overlap diagrams only need to be calculated
at the (usually small) proportion of pulse delays where they contribute.
We further show that heavy-tailed pulse envelopes, such as Lorentzians,
require calculation of overlap diagrams out to longer delay times
than Gaussian pulse envelopes.

We begin with an overview of perturbative spectroscopy calculations
in Sec.~\ref{sec:Spectroscopies-Overview} in order to establish
the construction of Feynman diagrams, with a unified perspective on
both phase matching and phase cycling spectroscopies. We describe
the algorithm of the DG in Sec.~\ref{sec:Diagram-Generator} including
optional methods to reduce the number of diagrams, depending on the
system and spectroscopy considered. We demonstrate the utility of
the DG by exploring EEI2D spectroscopy in Sec.~\ref{sec:Importance-of-overlap}.
We consider the same model system as in Ref.~\onlinecite{Suess2019},
and show that when optical pulses have slightly longer durations than
considered in that work, neither the impulsive limit nor the time-ordered
diagrams with pulse-shape effects included accurately predict spectra.
The calculation with all 240 diagrams requires less than twice the
time as using only the 7 time-ordered diagrams despite considering
34 times as many diagrams. These results underscore the importance
of including all of the additional 233 overlap diagrams, as well as
the utility of the DG in not only generating these diagrams, but also
automatically determining in which conditions they contribute.

\section{Nonlinear spectroscopy and Feynman diagrams \label{sec:Spectroscopies-Overview}}

We begin by establishing the standard perturbative framework of nonlinear
optical spectroscopy, from which Feynman diagrams are defined \citep{Mukamel1999}.
Consider a system with density matrix $\rho$, which evolves in the
absence of perturbation according to 
\[
\rho(t)=\mathcal{T}_{0}(t,t')\rho(t'),
\]
where $\mathcal{T}_{0}$ is a time evolution operator. We restrict
the following discussion to the case of Hamiltonian systems or Markovian
open systems, in which $\mathcal{T}_{0}(t,t')=\mathcal{T}_{0}(t-t')$,
which must be known or approximated in order to complete calculations,
but the resulting diagrams are broadly applicable to non-Markovian
situations, as well. For the purposes of diagram generation, however,
we simply need to assume $\mathcal{T}_{0}$ exists. We also define
the differential time evolution operator $\mathscr{L}_{0}$ so 
\[
\frac{d\rho(t)}{dt}=\mathscr{L}_{0}\rho(t).
\]

The perturbative optical fields are described as classical electric
fields $\mathbf{E}(t)$, which interact with the system in the electric-dipole
approximation through the perturbation Hamiltonian

\begin{equation}
H'(t)=-\cvec{\mu}\cdot\cvec E(t),\label{eq:Hprime-1}
\end{equation}
where $\cvec{\mu}$ is the electric dipole operator. Then the time
evolution of the system is 
\begin{equation}
\frac{d\rho(t)}{dt}=\mathscr{L}_{0}\rho(t)-\frac{i}{\hbar}\left[H'(t),\rho(t)\right].\label{eq:rhodot2}
\end{equation}
This form is the basis for diagramatic perturbation theory in $\mathbf{E}(t)$.

We describe $\mathbf{E}(t)$ as a sum over $L$ pulses, where each
pulse is denoted by a lowercase letter starting from $a$. A typical
$3^{rd}$-order signal is calculated using up to 4 pulses. We write
the electric field as 
\begin{equation}
\cvec E(t)=\sum_{j=a,b,...,L}\cvec e_{j}\varepsilon_{j}(t)+\cvec e_{j}^{*}\varepsilon_{j}^{*}(t)\label{eq:E(t)}
\end{equation}
where $\cvec e_{j}$ is the possibly complex polarization vector,
and the amplitude $\varepsilon_{j}$ of each pulse is defined with
envelope $A_{j}$, central frequency $\omega_{j}$, wavevector $\cvec k_{j}$,
and phase $\phi_{j}$ as 
\[
\varepsilon_{j}(t)=A_{j}(t-t_{j})e^{-i\left(\omega_{j}(t-t_{j})-\cvec k_{j}\cdot\cvec r-\phi_{j}\right)},
\]
where $t_{j}$ is the arrival time of each pulse. We make the physical
assumption that each pulse is nonzero in the finite interval $[t_{j,\text{min}},t_{j,\text{max}}]$.
The DG uses this range to determine when pulses overlap; the form
of $A_{j}(t)$ is unimportant for diagram generation. The light-matter
interaction is a sum over the rotating ($\varepsilon_{i}$) and counter-rotating
($\varepsilon_{i}^{*}$) terms. In the rotating wave approximation
(RWA), the rotating terms excite the ket-side and de-excite the bra-side
of the density matrix, respectively, and the counter-rotating terms
excite the bra-side and de-excite the ket side, respectively \citep{Mukamel1999}.

We treat the effect of the optical fields using standard time-dependent
perturbation theory and assume that at time $t_{0}$ the system is
in a stationary state of $\mathscr{L}_{0}$, which is $\rho^{(0)}$.
Then 
\begin{equation}
\rho(t)=\rho^{(0)}+\rho^{(1)}(t)+\rho^{(2)}(t)+...\label{eq:rho-expansion-1}
\end{equation}
where \citep{Mukamel1999}
\begin{equation}
\rho^{(n+1)}(t)=\int_{0}^{\infty}dt'\mathcal{T}_{0}(t')\left[\frac{-i}{\hbar}\cvec{\mu}\cdot\mathbf{E}(t-t'),\rho^{(n)}(t-t')\right].\label{eq:rhon_1}
\end{equation}
Using Eq.~\ref{eq:E(t)}, we define $\rho^{(n+1)}(t)$ as a sum over
four types of terms,
\begin{align}
\rho^{(n+1)}(t) & =\sum_{j}\left(K_{j}+K_{j^{*}}+B_{j}+B_{j^{*}}\right)\rho^{(n)}(t),\label{eq:KBdef}
\end{align}
where $K_{j}$ and $B_{j}$ are superoperators representing ket- and
bra-side actions, respectively, of the rotating terms of pulse $j$
on $\rho$, while $K_{j*}$ and $B_{j*}$ give the equivalent counter-rotating
terms. By inspection of Eq.~\ref{eq:rhon_1}, all four types of terms
in Eq.~\ref{eq:KBdef} can be compactly defined as 
\begin{equation}
O_{j^{(*)}}=\eta_{O}\frac{i}{\hbar}\int_{0}^{\infty}dt'\mathcal{T}_{0}(t')\left(\cvec{\mu}^{O}\cdot\cvec e_{j}^{(*)}\varepsilon_{j}^{(*)}(t-t')\right),\label{eq:Odef}
\end{equation}
where $O=K,B$, $\eta_{K}=1$ and $\eta_{B}=-1$, and we define dipole
superoperators $\mu^{K}\rho\equiv\mu\rho$ and $\mu^{B}\rho\equiv\rho\mu$.
The operators $\{O_{j^{(*)}}\}$ are the building blocks for all perturbative
spectroscopies. The full $\rho^{(n)}$ is constructed from the unperturbed
state $\rho^{(0)}$ as 
\begin{equation}
\rho^{(n)}(t)=\left[\sum_{j=a,b,\dots,L}\left(K_{j}+K_{j^{*}}+B_{j}+B_{j^{*}}\right)\right]^{n}\rho^{(0)},\label{eq:rhon_general}
\end{equation}
which involves $(4L)^{n}$ terms when the exponent and sum are fully
expanded. Each of these terms is a sequence of $n$ applications of
$O_{j^{(*)}}$ and can be represented as a Feynman diagram. Most of
these diagrams are unimportant for any given spectroscopy, and the
subset of diagrams that contributes to a particular experiment is
determined by the set of optical pulses and a phase-discrimination
condition.

Regardless of the computational details used to calculate the $\{O_{j^{(*)}}\}$,
all perturbative calculations can be represented and organized using
the same Feynman diagrams, and thus the DG is useful for any perturbative
spectroscopy algorithm. For example, UFSS contains two methods for
calculating the action of the $O_{j^{(*)}}$ operators, which are
derived for closed systems in Ref.~\onlinecite{Rose2019} and for
open systems in Ref.~\onlinecite{Rose2020b}. The integral form of
the $\{O_{j^{(*)}}\}$ in Eq.~\ref{eq:Odef} is convenient for $\alg$
and is the open-systems analogue of similar expressions derived for
wavefunctions \citep{Engel1991,Yuen-Zhou2014}.

The two widely used phase-discrimination conditions are phase matching
and phase cycling. Phase-matching conditions are achieved by using
pulses that travel along different directions denoted by wavevectors
$\cvec k_{j}$, converging to interact with a sample that is assumed
to be uniform over a volume large compared to the wavelength of the
pulses \citep{Mukamel1999}. The pulses induce a polarization field
in the sample $\mathbf{P}(t)=\langle\cvec{\mu}\rho(t)\rangle$, which
produces radiation in all directions. A detector placed in a direction
that satisfies the phase-matching condition $\cvec k_{d}=\sum_{j}m_{j}\cvec k_{j}$,
where $m_{j}$ are integers, is sensitive to a polarization field
that is described by only a subset of diagrams. Often one is interested
in the lowest-order signal in perturbation theory that contributes
to the given phase-discrimination condition. Heterodyne detection
with a local oscillator (LO) allows full determination of amplitude
and phase of the emitted radiation.

\begin{figure}
\includegraphics[width=0.4\paperwidth]{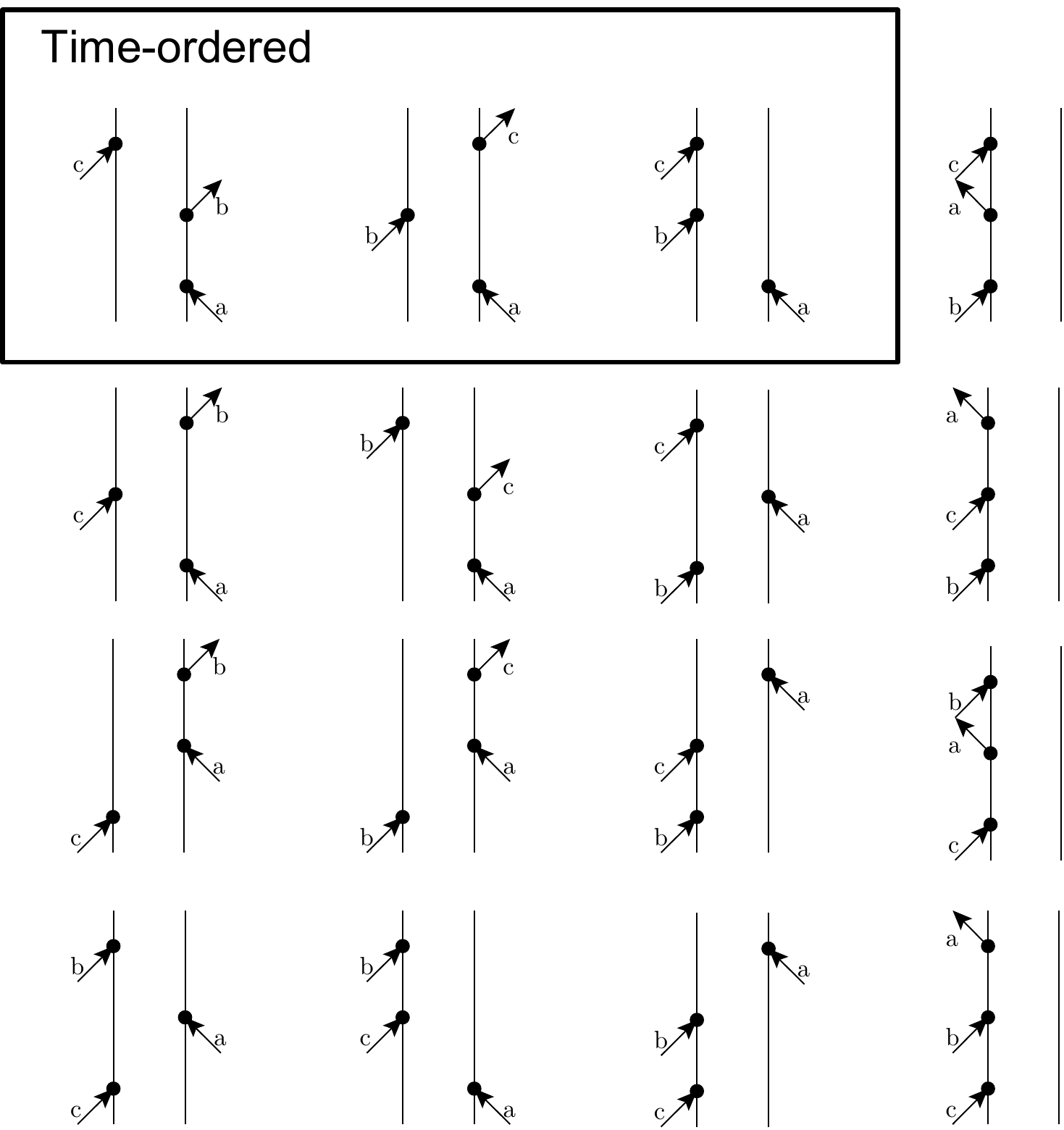}\caption{\label{fig:All-rephasing}All diagrams needed to calculate the rephasing
2DPE signal. The box encloses the 3 time-ordered diagrams that are
typically calculated. The other diagrams contribute to the signal
when two or more of the pulses overlap. These diagrams were automatically
generated and drawn using the diagram generator (DG). The DG can produce
both double-sided diagrams, as shown, and pairs of single-sided diagrams
for equivalent wavefunction calculations. }
\end{figure}

For example 2D photon echo (2DPE) signals involve three pulses, $a,b,c$,
that interact with the sample and a fourth LO pulse $d$. The 2DPE
rephasing and non-rephasing signals are measured with detectors placed
in the $\cvec k_{d}=-\cvec k_{a}+\cvec k_{b}+\cvec k_{c}$ (see Fig.~\ref{fig:All-rephasing})
and $\cvec k_{d}=\cvec k_{a}-\cvec k_{b}+\cvec k_{c}$ directions,
respectively. These signals are calculated using 
\begin{align*}
\cvec P_{\mathrm{k}_{d}}^{(3)}(t) & =\langle\cvec{\mu}\rho_{\mathrm{k}_{d}}^{(3)}(t)\rangle\\
\tilde{\cvec P}_{\mathrm{k}_{d}}^{(3)}(\omega) & =\int_{-\infty}^{\infty}dte^{i\omega t}\cvec P_{\mathrm{k}_{d}}^{(3)}(t)\\
S_{\mathrm{k}_{d}}^{(3)}(\omega) & =\mathrm{Im}\left[\tilde{\varepsilon}_{d}^{*}(\omega)\cvec e_{d}\cdot\tilde{\cvec P}_{\mathrm{k}_{d}}^{(3)}(\omega)\right]
\end{align*}
where $S_{\mathrm{k}_{d}}^{(3)}(\omega)$ are the signals and $\rho_{\mathrm{k}_{d}}^{(3)}(t)$
is the portion of $\rho^{(3)}(t)$ that produces radiation in the
$\mathbf{k}_{d}$ direction; the primary purpose of diagrammatic perturbation
theory is to organize the efficient calculation of $\rho_{\mathrm{k}_{d}}^{(3)}(t)$
without needing to calculate all contributions to $\rho^{(3)}(t)$.

An alternative phase discrimination method uses phase cycling over
the relative phases of collinear pulses. This method generally detects
a signal proportional to an excited state population, such as fluorescence
or photocurrent \citep{Tekavec2007,Perdomo-Ortiz2012,Nardin2013,Bakulin2016}.
A fourth-order signal $S_{d}^{(4)}(t)$ in such a setup is calculated
as 
\[
S_{d}^{(4)}(t)=\left\langle Q\rho_{d}^{(4)}(t)\right\rangle ,
\]
where $Q$ is a projection operator onto the relevant excited electronic
states and $\rho_{d}^{(4)}(t)$ includes only those contributions
to $\rho^{(4)}(t)$ that contribute to the chosen phase-cycling condition.

For example, the 2DPE rephasing signal is composed of the 16 diagrams
in Fig.~\ref{fig:All-rephasing}, which were automatically generated
and drawn using the DG. If, however, $t_{b,\text{min}}>t_{a,\text{max}}$
and $t_{c,\text{min}}>t_{b,\text{max}}$ (that is, the pulses are
time ordered), then only three diagrams contribute, shown in the black
box.

The operators $\{O_{j^{(*)}}\}$ as an abstract concept have previously
been introduced in various forms \citep{Engel1991,Yuen-Zhou2014}.
To our knowledge, they have not been used for the purposes of automated
calculations before. In Sec.~\ref{sec:Diagram-Generator}, we describe
how users inputs the pulse intervals and phase discrimination conditions
and present the algorithm that produces the relevant diagrams. In
Sec.~\ref{sec:Importance-of-overlap} we demonstrate the DG by using
up to 240 diagrams and show the importance of including the full set
of diagrams when performing simulations with finite pulse durations.

\section{Diagram Generator\label{sec:Diagram-Generator}}

The user of the DG inputs the desired phase discrimination condition
and pulse intervals, and the DG automatically generates all Feynman
diagrams. The DG is much less computationally expensive than the evaluation
of the contribution from each diagram, so it is intended to be rerun
for each set of desired pulse delays. The DG then returns pulse-overlap
diagrams only when the pulse overlaps allow them to contribute. This
package generates diagrams as a list of the operators $\{O_{j^{(*)}}\}$
defined in Eq.~\ref{eq:Odef}. The $\alg$ and RKE tools outlined
in Ref.~\onlinecite{Rose2020b} and included in UFSS are designed
to interpret this list and produce spectra. Alternatively, the generated
diagrams could be exported to another calculation engine. The DG can
also draw them for inspection.

Feynman diagrams are determined by the number of pulses and the phase-discrimination
condition. Phase-discrimination determines the number of interactions
with the rotating or counter-rotating part of each pulse that contribute
to the measured signal. The user inputs the number of each type of
interaction as a list of tuples $[(n_{r}^{a},n_{c}^{a}),(n_{r}^{b},n_{c}^{b}),...,(n_{r}^{L},n_{c}^{L})]$.
This list corresponds to detection with $\mathbf{k}_{d}=\sum_{j=a,b,...,L}(n_{r}^{j}-n_{c}^{j})\mathbf{k}_{j}$
in the case of phase matching and to detection of signals in phase
with $\sum_{j=a,b,...,L}(n_{r}^{j}-n_{c}^{j})\phi_{j}$, where $\phi_{j}$
are the modulated phases, in the case of phase cycling \citep{Tekavec2007}.
The order of spectroscopy is given by $n_{\text{total}}=\sum_{j=a,b,...,L}n_{r}^{j}+n_{c}^{j}$.
In order to determine which diagrams are causally allowed, the user
must also input the time intervals $[t_{j,\text{min}},t_{j,\text{max}}]$
when each pulse is nonzero as a list. The user updates this list for
each set of pulse delay times desired. The number of diagrams that
must be considered when dealing with finite pulses expands dramatically
as one considers higher-order signals, such as exciton-exciton interaction
2D spectra (the 3-pulse $5^{th}$-order signal measured in the $\cvec k_{d}=-2\mathbf{k}_{1}+2\mathbf{k}_{2}+\mathbf{k}_{3}$
direction, see Sec~\ref{sec:Importance-of-overlap}), where the number
of time-ordered diagrams is 7 \citep{Suess2019}. However, when all
pulses overlap, there are 240 diagrams.

\begin{figure*}
\includegraphics[width=1\textwidth]{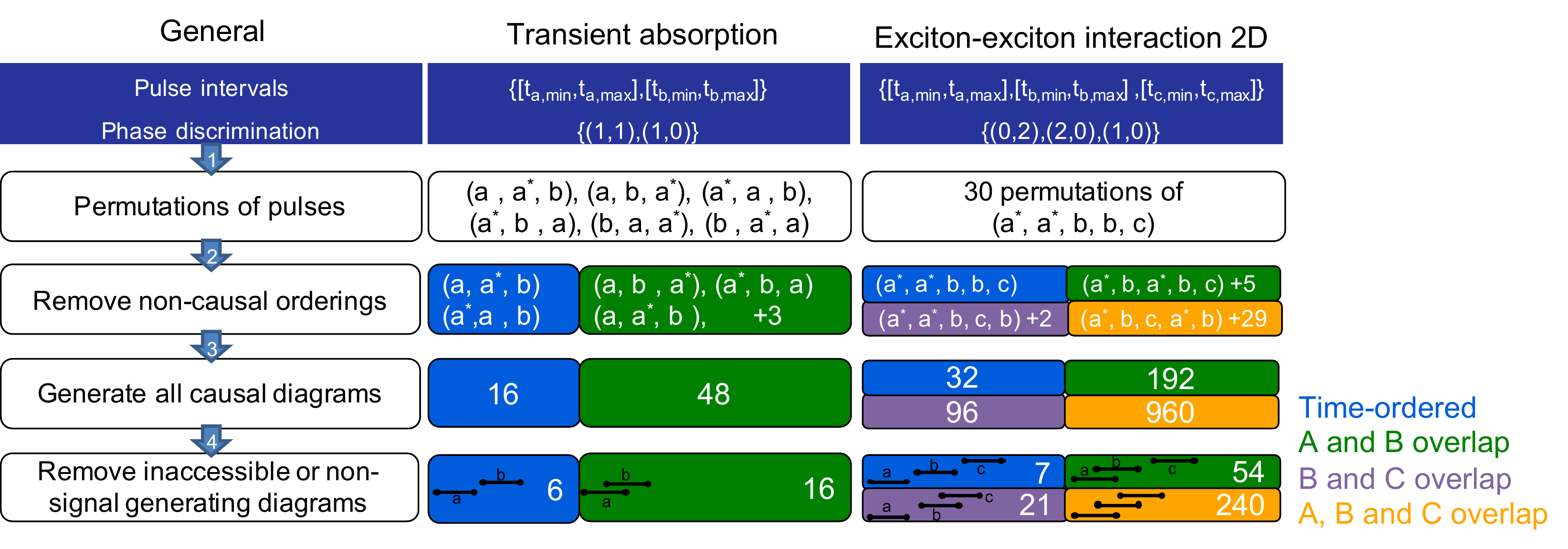}

\caption{\label{fig:DG-figure}Graphical outline of automated diagram generation.
Left column is the general case, while the center and right columns
show specific examples. User inputs of pulse intervals and phase discrimination
of the desired experiment are in the top box. The steps of the algorithm,
numbered in the arrows, are described in the main text. After step
3, the numbers in the boxes indicate the number of diagrams at that
stage. Blue boxes show results with non-overlapping pulses. Green
shows results with pulses $a$ and $b$ overlapping, purple shows
results with pulses $b$ and $c$ overlapping, and gold shows results
with pulses $a$, $b$ and $c$ all overlapping. Inset black bars
in the bottom row graphically indicate these pulse overlaps. }
\end{figure*}

We outline the steps for determining all diagrams that contribute
to the signal for a given pulse configuration in Fig.~\ref{fig:DG-figure},
which includes the general scheme and two examples.

(1) Starting from $[(n_{r}^{i},n_{c}^{i})]_{i=a,b,...,L}$, we begin
with a canonical list of interactions, where we list the pulses in
order of pulse number starting from $a$, with rotating terms coming
before counter-rotating terms (see Fig.~\ref{fig:DG-figure} for
examples). The length of the resulting list is $n_{\text{total}}$.
For each pulse there are $n_{r}^{j}$ repetitions of ``$j$'' with
the rotating term and $n_{c}^{j}$ repetitions of ``$j^{*}$'' with
the counter-rotating term, as shown in Fig.~\ref{fig:DG-figure}.
We then generate all unique permutations of this canonical list of
interactions. The number of unique permutations is 
\[
\frac{n_{\text{total}}!}{\Pi_{i}n_{r}^{i}!n_{c}^{i}!}.
\]

(2) Using the list of pulse intervals $\{[t_{j,\text{min}},t_{j,\text{max}}]\}_{j=a,b,...L}$,
non-causal orderings are removed. For each permutation of the time-ordered
list, we check whether $t_{\text{min}}$ of each pulse occurs before
$t_{\text{max}}$ of each following pulse in the list and remove the
permutation if not.

(3) For each permutation from step 2, we generate all of the allowed
diagrams that satisfy the phase-discrimination conditions. Each interaction
can occur either on the ket-side ($K_{j^{(*)}}$) or the bra-side
($B_{j^{(*)}}$), giving $2^{n_{\text{total}}}$ diagrams associated
with each permutation from step 2. For example, the interaction $a^{*}$
can act as $K_{a^{*}}$ or as $B_{a^{*}}$, while the interaction
$b$ can act as either $K_{b}$ or $B_{b}$.

At this point we have the maximum number of diagrams that could contribute
to the calculation, given the phase-discrimination condition and pulse
intervals. However, many of these diagrams do not contribute under
common assumptions. Using some minimal information about the material
system that is being modeled, many of these diagrams can be removed
in the optional step 4.

(4) (Optional)
\global\long\def\labelenumi{(\alph{enumi})}%
\begin{enumerate}
\item Keep only diagrams that remain in accessible states. For instance,
consider optical spectroscopy of a system that has three optically
separated manifolds, each separated by an energy gap $E_{g}$. If
the temperature is much less than $E_{g}$, we can approximate that
the initial thermal state is entirely in the ground-state manifold.
Any diagram that includes excitation above the highest manifold or
de-excitation from the lowest manifold is removed. We track the number
of optical excitations by assigning manifold indices for both the
ket and bra sides of a density matrix. We say that the initial density
matrix $\rho^{(0)}$ is entirely composed of ground-state populations,
and therefore it has manifold indices $[0,0].$ We then assign the
following rules describing the action of the $O_{j^{(*)}}$ operators:
\begin{align*}
K_{j} & :[+1,0]\\
K_{j^{*}} & :[-1,0]\\
B_{j} & :[0,-1]\\
B_{j^{*}} & :[0,+1]
\end{align*}
We apply these rules in succession for a diagram and track the indices
$[i,l]$ after each interaction. If either $i$ or $l$ drop below
0 or rise above the maximum allowed manifold, the diagram is removed.
When the manifolds are coupled by relaxation processes, we no longer
remove diagrams that rise above the allowed maximum manifold, since
population can decay to a lower manifold and then be excited up again
by a subsequent interaction. We do still remove diagrams where $i$
or $l$ drop below zero. 
\item The integer logic of part 4(a) is also helpful in determining which
diagrams contribute to the final signal. For spectroscopies that measure
the emitted polarization field, we are interested in the object $\Tr[\mu\rho]$.
Typically in optical spectroscopy, $\mu$ connects only adjacent manifolds
(either because this is an accurate model for the dipole operator
or because the measurement bandwidth only supports 1-manifold transitions).
The components of $\rho$ that contribute are then coherences between
adjacent optical manifolds. Therefore, we filter out all diagrams
except those that end in a state $[i+1,i]$ (Note that the diagrams
that end in $[i,i+1]$ are physically valid; however, we do not calculate
them, as they are the Hermitian conjugate pairs of the calculated
diagrams \citep{Mukamel1999}). Note that we again cannot apply this
filter when $\mathcal{T}$ includes inter-manifold relaxation. 
\item If the final observable is instead linked to excited-state populations,
as in the case of fluorescence or photo-current detection, we look
only for diagrams that end in a population $[i,i]$ where $i>0$.
Again, we cannot apply this filter when $\mathcal{T}$ includes inter-manifold
relaxation.
\end{enumerate}
Each of the diagram reductions from step 4 can be turned off with
a flag or modified to suit an accurate model for the system in question.
For example, in vibrational spectroscopy, rule 4a) does not apply,
but rules 4b) or 4c) (or modifications of them) may still apply. We
include a variety of examples of step (4) in the Jupyter notebook
$\texttt{DiagramGeneratorExample.ipynb}$, including an example of
infrared vibrational transient absorption that cannot remove any of
the causal diagrams and must use all 16 (48 for overlapping pulses)
diagrams that come from step (3).

Thus far we have described how the DG creates the double-sided Feynman
diagrams associated with density-matrix-based calculations. The DG
can also create the one-sided Feynman diagrams used for wavefunction-based
calculations \citep{Mukamel1999,Yuen-Zhou2014}. This procedure begins
by creating all relevant double-sided diagrams. Each double-sided
diagram is converted to a pair of one-sided diagrams associated with
the bra and ket wavefunctions. Since wavefunction-based calculations
do not impose time-ordering between the bra and ket sides \citep{Mukamel1999},
several of the double-sided diagrams map onto the same pair of single-sided
diagrams. After creating all of the pairs of single-sided diagrams,
we eliminate duplicates to avoid over-counting some pathways.

The DG both produces a full set of contributing diagrams to be calculated
and updates that list as the pulse timings change. In the example
$3$-pulse $5^{th}$-order spectroscopy considered in Sec.~\ref{sec:Importance-of-overlap},
the pulse delays vary to produce frequency-domain signals. Only a
small number of pulse configurations have all three pulses overlapping.
In those cases, 240 diagrams must be calculated, but as the pulse
delays change, calculations can be reduced to 54, 21, or 7 diagrams,
which all occurs automatically. Including overlap diagrams increases
the number of required diagrams by a factor of over 30, but since
diagrams are only included when pulse delays require, calculations
of frequency-domain spectra are only 1.5-5 times longer than those
including only the 7 time-ordered diagrams, depending upon which pulse
delays are calculated.

\section{Importance of overlap diagrams\label{sec:Importance-of-overlap}}

\begin{figure}
\includegraphics[width=0.4\paperwidth]{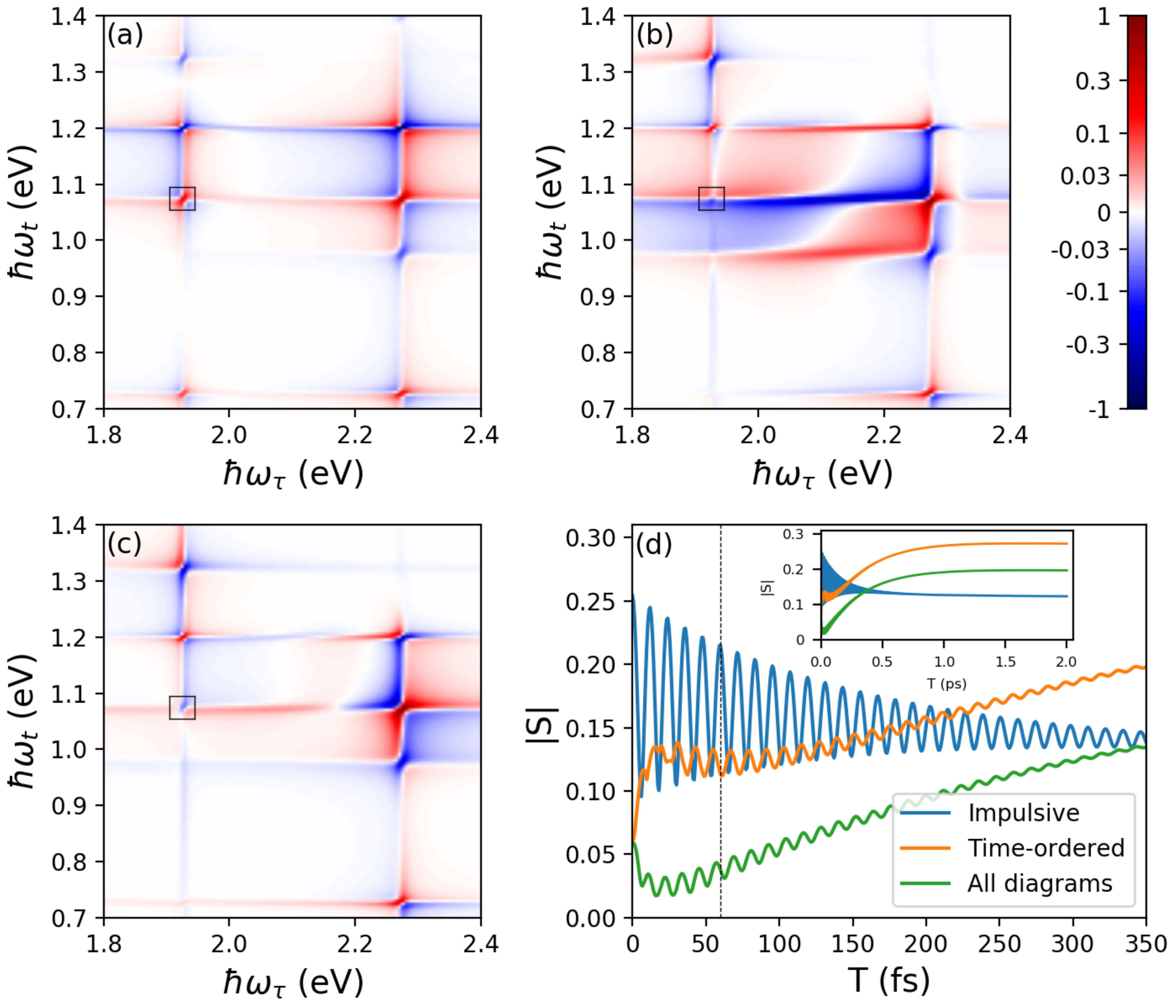}\caption{\label{fig:Engel-comparison} (a)-(c) Real part of the EEI2D spectrum
at a delay time of $T=60\,\text{fs}$ for a dimer of three-level systems
with a relaxation rate of 0.015 $\text{fs}^{-1}$, as described in
the text, with different electric field shapes. Horizontal axes show
the Fourier transform of the delay time $\tau$ between the first
and second pulses. Vertical axes show the detection frequency. EEI2D
spectra were calculated for $275\times275$ values of $\tau,t$ and
then Fourier transformed. Note the signed-logarithmic color scale.
(a) Impulsive calculation, which requires only time-ordered diagrams,
(b) 15~fs FWHM Gaussian pulses using only time-ordered diagrams,
(c) 15~fs FWHM Gaussian pulses including all 54 diagrams that contribute
at this $T$. (a)-(c) are normalized independently so that the largest
peak has magnitude 1. (d) Maximum of the absolute value of the signal
contained within the black boxes in (a)-(c) as a function of delay
time $T$. Dashed line shows $T$ used in (a-c). The inset shows a
longer time window and the steady-state that is reached by 1~ps.
The impulsive and time-ordered calculations both quantitatively and
qualitatively deviate from the full calculation.}
\end{figure}

Spectra are often calculated assuming impulsive pulses. Even when
finite pulse shapes are considered, the overlap diagrams are often
neglected, as in the only two studies of finite pulses in $5^{th}$-order
spectroscopies of which we are aware \citep{Do2018,Suess2020}. In
some cases these approximations are warranted, but sometimes the overlap
diagrams are important to understand signals both quantitatively and
qualitatively. Here, we demonstrate an example where neglecting overlap
diagrams leads to significant artifacts in predicted spectra.

We consider a model system used in Ref.~\onlinecite{Suess2019} to
study exciton-exciton annihilation, using three-pulse exciton-exciton
interaction 2D (EEI2D) spectroscopy. The model consists of a dimer
of two coupled three-level systems (3LS). The Hamiltonian takes the
form 
\begin{align*}
H_{0} & =E_{gg}\left(\ket{gg}\bra{gg}\right)\\
 & +E_{eg}\left(\ket{eg}\bra{eg}+\ket{ge}\bra{ge}\right)+J\left(\ket{eg}\bra{ge}+h.c.\right)\\
 & +E_{ee}\left(\ket{ee}\bra{ee}\right)+E_{fg}\left(\ket{fg}\bra{fg}+\ket{gf}\bra{gf}\right)\\
 & +K\left(\ket{fg}\bra{ee}+\ket{gf}\bra{ee}+h.c.\right)\\
 & +E_{ef}\left(\ket{fe}\bra{fe}+\ket{ef}\bra{ef}\right)+L\left(\ket{fe}\bra{ef}+h.c.\right)\\
 & +E_{ff}\left(\ket{ff}\bra{ff}\right)
\end{align*}
where $E_{ij}=E_{i}+E_{j}$, all of the constants are defined in Table~\ref{tab:BrixnerEngel},
$\ket g,\ket e,\ket f$ are the ground, singly excited, and doubly
excited states, respectively, of each 3LS, and $\ket{uv}=\ket u_{1}\otimes\ket v_{2}$.
Following Ref.~\onlinecite{Suess2019}, we consider relaxation processes
at zero temperature for each isolated monomer unit from $\ket f_{i}$
to $\ket e_{i}$ at rate $k_{M}$ and neglect relaxation from $\ket e_{i}$
to $\ket g_{i}$, since it is not important for exciton-exciton interactions.
Reference \onlinecite{Suess2019} included relaxation using the stochastic
Schrodinger equation, while we include relaxation using the Lindblad
formalism with
\[
\dot{\rho}=-\frac{i}{\hbar}[H_{0},\rho]-\sum_{n\neq m}k_{nm}L\left[\ket m\bra n\right]\rho
\]
where $\ket m$ and $\ket n$ are eigenstates of $H_{0}$ and the
Lindblad superoperator $L[O]$ is defined by
\[
L[O]\rho=2O\rho O^{\dagger}-O^{\dagger}O\rho-\rho O^{\dagger}O.
\]
We follow Ref.~\onlinecite{Suess2019} by projecting the monomer
relaxation rates into the eigenstates $\ket n$ with energy $E_{n}$.
New relaxation rates coupling the eigenstates of $H_{0}$ are defined
as 
\begin{align*}
k_{nm}= & \bigg(\big|\braket n{fg}\big|^{2}\big|\braket{eg}m\big|^{2}+\big|\braket n{gf}\big|^{2}\big|\braket{ge}m\big|^{2}\\
 & +\big|\braket n{fe}\big|^{2}\big|\braket{ee}m\big|^{2}+\big|\braket n{ef}\big|^{2}\big|\braket{ee}m\big|^{2}\bigg)k_{M}.
\end{align*}
The resulting rates range from 0.0075~fs$^{-1}$ to 0.0016~fs$^{-1}$
where the smallest rate corresponds to decay from the lowest-energy
double-exciton state to the lowest-energy single-exciton state.

EEI2D is designed to probe the dynamics of the doubly excited states.
Diagonalizing $H_{0}$ shows that the two optically bright doubly
excited states are separated by $0.35\,\textrm{eV}$, corresponding
to an oscillation period of $T_{o}=12\,\textrm{fs}$, which is the
fastest important oscillation in this system. One generally assumes
that pulses significantly shorter than this period will be well-approximated
by impulsive pulses but that pulses that interact with the system
on a comparable timescale require more careful treatment. In this
section we study the effects of using Gaussian pulses with a full-width
half-maximum (FWHM) of $15\,\textrm{fs}$, with a comparison to Lorentzian
pulses at the end.

The perturbative calculations of optical signals were performed using
the Ultrafast Ultrafast ($\alg$) Spectroscopy method detailed in
Ref.~\onlinecite{Rose2020b}. $\alg$ calculates the fifth-order
polarization signal as a function of the delay time $\tau$ between
pulses $a$ and $b$ and the delay time $T$ between pulses $b$ and
$c$ as
\begin{align*}
\cvec P_{\mathrm{k}_{d}}^{(5)}(\tau,T,t) & =i\langle\cvec{\mu}\rho_{\mathrm{k}_{d}}^{(5)}(t)\rangle,
\end{align*}
where $t$ is the time measured after the arrival of pulse $c$. The
2D frequency-frequency correlation spectrum is calculated as
\[
\tilde{\cvec P}_{\mathrm{k}_{d}}^{(5)}(\omega_{\tau},T,\omega_{t})=\frac{1}{2\pi}\int_{-\infty}^{\infty}d\tau e^{-i\omega_{\tau}\tau}\int_{-\infty}^{\infty}dte^{i\omega_{t}t}\cvec P_{\mathrm{k}_{d}}^{(5)}(\tau,T,t),
\]
which is approximated using the discrete Fourier transform. We use
$\tilde{\cvec P}_{\mathrm{k}_{d}}^{(5)}(\omega_{\tau},T,\omega_{t})$
as a proxy for the signal field, which would in practice be detected
using heterodyne detection with a local oscillator.

\begin{table}
\caption{\label{tab:BrixnerEngel}Values used in the Hamiltonian $H_{0}$ and
bath coupling rates for the model dimer system studied using EEI2D,
adapted from Ref.~\onlinecite{Suess2019}.}

\begin{tabular}{cccc}
 &  &  & \tabularnewline
\hline 
\hline 
 & $E_{g}$  & $E_{e}$  & $E_{f}$\tabularnewline
Energy (eV)  & 0.0  & 1.0  & 2.2\tabularnewline
\hline 
 & $J$  & $K$  & $L$\tabularnewline
Coupling (eV)  & 0.2  & 0.1  & 0.05\tabularnewline
\hline 
 & $k_{M}$  &  & \tabularnewline
Rate ($\text{fs}^{-1}$)  & 0.015  &  & \tabularnewline
\hline 
\hline 
 &  &  & \tabularnewline
\end{tabular}
\end{table}

We compare calculations of EEI2D spectra using three approximations:
impulsive pulses, finite pulses with only time-ordered diagrams, and
finite pulses including all diagrams. Figure \ref{fig:Engel-comparison}(a)
shows the impulsive limit with a delay time of $T=60\,\text{fs}$.
In keeping with Ref.~\onlinecite{Suess2019}, we perform all calculations
with $\tau$ and $t$ each ranging from $0$ to $823\,\text{fs}$,
and multiply the $\tau$ and $t$ axes with a Gaussian window function
of $\sigma=200\,\text{fs}$ to avoid ringing effects from the discrete
Fourier transform. We calculate spectra for 275 values of $\tau$
and $t$, which are more than sufficient to produce well-resolved
frequency-frequency spectra. Reference \onlinecite{Suess2019} performs
calculations in the impulsive limit and states that calculations using
time-ordered diagrams with 5 fs FWHM pulses are visually nearly identical
to the impulsive limit, which we also find. Finite pulse effects are,
unsurprisingly, unimportant when the pulse durations are shorter than
$T_{o}$, the fastest timescale in the system. We use this model system
to illustrate the differences that occur with only modestly longer
pulses.

We consider Gaussian pulses with FWHM of 15 fs, similar to $T_{o}$.
Figure \ref{fig:Engel-comparison}(b) shows calculations using finite
pulses but only the 7 time-ordered diagrams, showing clear differences
from the impulsive limit in \ref{fig:Engel-comparison}(a). Figure
\ref{fig:Engel-comparison}(c) shows results using the same finite
pulses but including all of the 233 additional pulse overlap diagrams
at delay times when they are required. The visual difference between
Fig.~\ref{fig:Engel-comparison}(b) and (c) demonstrates the importance
of including pulse-overlap diagrams in addition to finite pulse effects
in time-ordered diagrams. Simply adding finite pulse effects to the
time-ordered diagrams is not sufficient for making good spectroscopic
predictions.

Figure \ref{fig:Engel-comparison}(d) shows the magnitude of the cross
peak contained in the solid box in panels (a)-(c), and the oscillation
period $T_{o}$ is clearly visible. The oscillations at that peak
correspond to absorption into one and emission from another doubly
excited eigenstate of $H_{0}$. This peak is clearly visible in the
impulsive limit in Fig.~\ref{fig:Engel-comparison}(a), but is dominated
by significant horizontal streaks in Fig.~\ref{fig:Engel-comparison}(b),
which bleed over from the stronger peak at $\hbar\omega_{\tau}=2.3$~eV.
This significant extension of the peaks in the $\omega_{\tau}$ direction
is an artifact of neglecting the overlap diagrams, demonstrated by
its removal in Fig.~\ref{fig:Engel-comparison}(c). Figure \ref{fig:Engel-comparison}(d)
shows that both the visibility of the oscillation and the overall
evolution of the envelope are qualitatively different in the three
studied approximations. We vary $T$ from 0 to 2 ps, for a total of
2667 different values of $T$. At long $T$ (shown in the inset),
the doubly excited states decay and there is a static excited state
absorption signal from the singly excited population, with the timescale
of saturation and oscillation decay matching the slowest $k_{nm}$.
Neither impulsive nor time-ordered calculations with finite pulses
accurately predict this EEI2D spectrum despite using an ultrafast
optical pulse.

Note that the differences between the full calculation and the one
using only the time-ordered diagrams persist even with delay times
$T$ much greater than the pulse durations. In constructing $\tilde{\cvec P}_{\mathrm{k}_{d}}^{(5)}(\omega_{\tau},T,\omega_{t})$,
contributions with $\tau$, $t$ smaller than the pulse duration are
always included, making pulse-overlap effects apparent even at long
$T$ and requiring calculation of the overlap diagrams. Since the
DG only produces the extra diagrams for time delays that merit their
evaluation, the full calculation is not much more expensive than the
case with 7 time-ordered diagrams. For example, the calculation of
the green and orange curves in Fig.~\ref{fig:Engel-comparison}(d),
out to 350 fs, took 33 and 17 minutes on a 2017 MacBook Pro, respectively,
and included over 120,000 combinations of $\text{\ensuremath{\tau}}$
and $T$. The equivalent curves out to 2~ps, as in the inset, required
155 and 93 minutes with the same density of $T$ points; those involved
over 700,000 pulse delay combinations. Even though the full calculation
includes contributions from 34 times more diagrams, it required less
than twice the time to run than the time-ordered calculation.

\begin{figure}
\includegraphics[width=0.43\paperwidth]{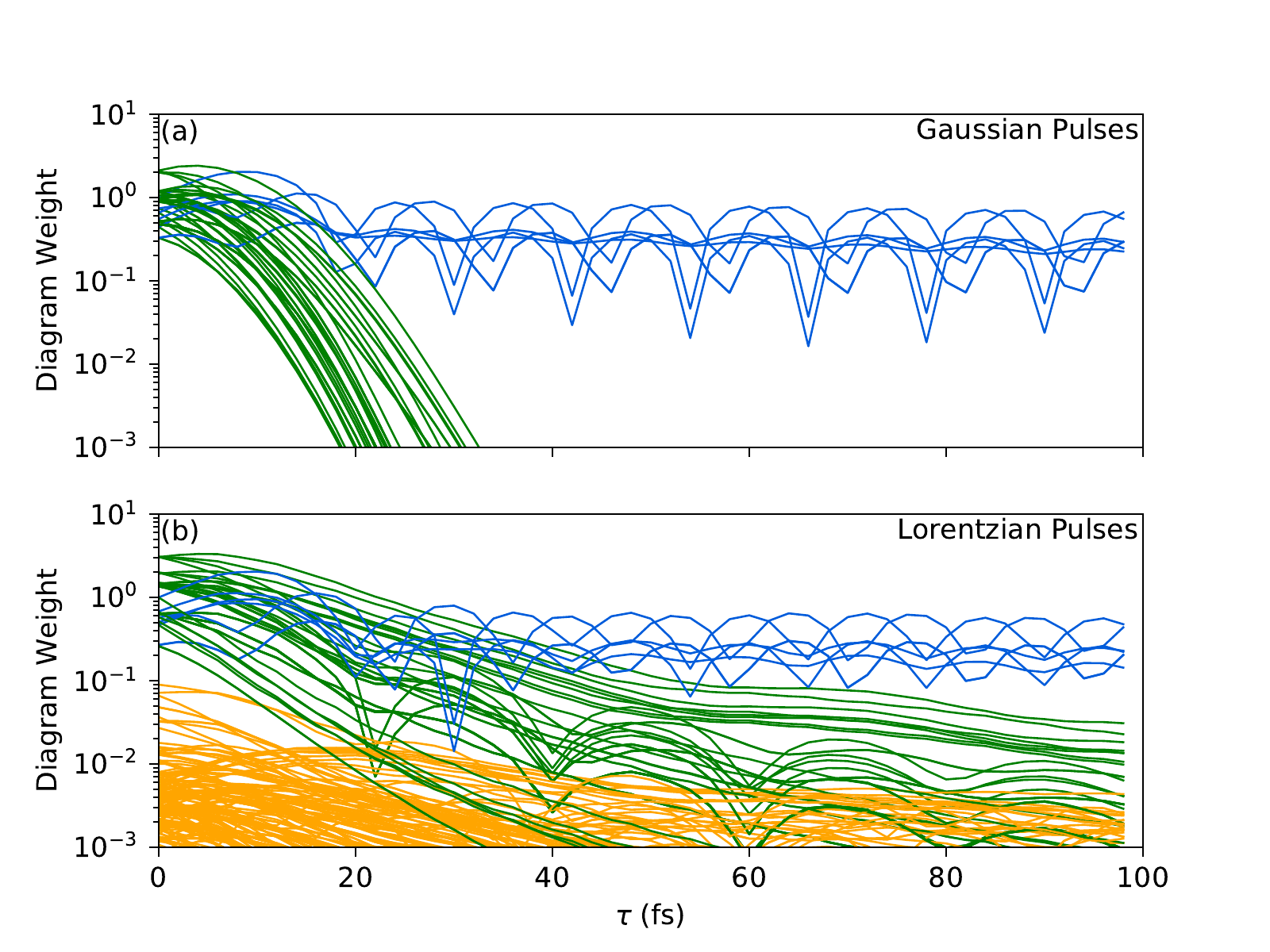}\caption{\label{fig:GaussLorentzDiagramWeights} Weight of all 240 EEI2D diagrams
for the same system as in Fig.~\ref{fig:Engel-comparison} at $T=100\,\text{fs}$
as a function of $\tau$ for 15~fs FWHM pulses with (a) Gaussian
and (b) Lorentzian time-domain envelopes. Blue lines show the weight
of the 7 time-ordered diagrams. Green lines show the weight of the
47 additional diagrams that contribute when pulses $a$ and $b$ overlap.
Panel (b) also shows the weights of 176 diagrams that contribute only
when all three pulses overlap (gold), whose weights are too small
to appear in (a); 10 more orange lines have weights less than $10^{-3}$.
Diagram weights are normalized so that the strongest time-ordered
diagram has a weight of 1 at $\tau=0$. Line colors for each set of
diagrams match those in Figure \ref{fig:DG-figure}.}

\end{figure}

The DG determines which diagrams contribute based solely upon the
user-chosen intervals where the pulses are nonzero. For well-behaved
pulses such as Gaussians with standard deviation $\sigma$, convergence
of spectra within 1\% is obtained for the results in Fig.~\ref{fig:Engel-comparison}
when the pulses are declared to be zero after $4\sigma$. However,
for pulses with heavy tails, extra care must be taken to determine
the correct pulse interval. We compare Gaussian pulses, as used in
Fig.~\ref{fig:Engel-comparison}, to Lorentzian pulses, which have
heavy tails. Figure~\ref{fig:GaussLorentzDiagramWeights} shows the
weight $D\int d\omega_{t}\left|\tilde{\cvec P}_{d}^{(5)}(\tau,T,\omega_{t})\right|$
for each of the 240 diagrams $d$ as a function of $\tau$ at $T=100\,\text{fs}$
for the system in Fig.~\ref{fig:Engel-comparison}, where $\tilde{\cvec P}_{d}^{(5)}$
is the signal due to a single diagram and $D$ is chosen so that at
$\tau=0$, the largest weight of a time-ordered diagram is 1. We consider
Gaussian and Lorentzian pulse envelopes with the same FWHM of 15~fs.

We observe that for both sets of pulses, at $\tau=0$ all 47 $a$,
$b$ overlap diagrams (green) have similar weight and therefore must
be calculated, and the same is true of all 233 overlap diagrams at
$\tau=T=0$ (not shown). With Gaussian pulses, all of the overlap
diagrams decay rapidly when $\tau$ exceeds the nominal pulse duration
and are negligible when $\tau>30$~fs. In contrast, with Lorentzian
pulses of the same nominal duration, even some diagrams where pulse
$c$ arrives before pulse $a$ or $b$ (shown in gold) are surprisingly
close in weight to the time-ordered diagrams at $\tau=0$, even though
the center of pulse $c$ arrives 100~fs after the first two. Further,
some of the $a$,$b$ overlap diagrams (green) continue to have weight
equal to time-ordered diagrams (blue) for $\tau$ approaching $60\,\text{fs}$,
four times the pulse FWHM; given the large number of overlap diagrams,
their contributions must be calculated even for considerably larger
$\tau$. It is clear that in the case of Lorentzian or other heavy-tailed
pulse envelopes, care must be taken in choosing the pulse intervals
for use with the DG, and users are encouraged to ensure that their
results are converged. Choosing a large pulse interval ensures accuracy
but increases computational cost as more overlap diagram contributions
must be calculated.

Many ultrashort optical pulses have heavy tails in the time domain,
and Ref.~\onlinecite{Palecek2019} showed experimental evidence that
pulse overlap effects can extend for up to 100~fs when using 17~fs
FWHM pulses in 2DPE experiments. They attributed this effect to the
significant wings on the pulse envelopes, which is consistent with
our analysis of Lorentzian pulses.

\section{Conclusion}

We have presented the diagram generator (DG), a tool for automatically
generating the Feynman diagrams that contribute to perturbative nonlinear
optical spectroscopies. The DG automatically determines when pulses
overlap and only generates extra overlap diagrams when required. This
automated process allows users to get the full advantage of including
all causally allowed diagrams with a low computational cost.

We have shown that including these overlap diagrams can be important
to correctly predicting or interpreting spectra when pulses are not
in the impulsive limit. Using EEI2D as an example, we have shown significant
errors when using only the time-ordered diagrams with pulses whose
duration is similar to the dynamics of the system. We have also shown
that overlap diagrams can make significant contributions to the signal
for delay times that are large relative to the pulse durations, when
those pulses have heavy temporal tails, as in the case of Lorentzian
pulses.

The DG is one module of the larger Ultrafast Spectroscopy Suite. The
other components in UFSS are described in Refs.~\citep{Rose2019,Rose2020b}.
Taken together, UFSS is a tool for automatically calculating arbitrary-order
spectroscopic signals while accounting for effects of finite pulse
shapes, which can be of arbitrary form. The diagrams produced by the
DG can also be used in other analytical or numerical tools. The DG
can be used for determining all of the diagrams that contribute to
any order spectra, whether in the impulsive limit or with finite pulses.
The DG may open the door to more easily calculating higher-order corrections
to commonly used $3^{\text{rd}}$-order spectroscopies and may lend
itself to developing intuition for higher-order spectroscopic techniques.

\begin{acknowledgments}
We acknowledge support from the Natural Sciences and Engineering Research
Council of Canada (NSERC) and the Ontario Trillium Scholarship.
\end{acknowledgments}

\subsection*{Data Availability Statement}

The data that support the findings of this study are available from
the corresponding author upon reasonable request. In addition, the
code to generate all of the figures in this manuscript is available
at \href{https://github.com/peterarose/ufss}{github}.

\bibliographystyle{aipnum4-1}
\bibliography{OpenUF2Paper}

\end{document}